# Structural stability, electronic structure and optical properties of dimension controlled self-assembled structures from clusters of cadmium telluride

Kashinath T Chavan, Sharat Chandra
*Materials Science Group, Indira Gandhi Centre for Atomic Research, Kalpakkam 603102, Tamil Nadu, India*
*Homi Bhabha National Institute, Training School Complex, Anushakti Nagar, Mumbai, 400094, India*

## Abstract

We report the first principle theory-based study of stability, electronic structure and optical properties of cluster assembled materials in various 1D, 2D and 3D nanostructures using a cage-like $Cd_9Te_9$ cluster as the super-atom. The bulk 3D self-assemblies form in 2D stacked structures for different cubic lattices. The face centered stacking is the most stable as compared to the simple cubic, body centered and zinc blende type stackings. The 2D stacks are formed as cluster assembled monolayers and the monolayer derived from the face centered structure is most stable. Further, the cluster chains (or wires) with more number of inter-cluster bonds are also seen to be dynamically stable. The electronic structure, bandgap, dielectric constant and absorption spectra along with the phonon dispersions are discussed for these self-assembled nanostructures.

## Keywords

## Introduction

The recent developments in nanotechnology significantly benefit from the bottom-up approach complementing the top-down approach, since the former is a high-throughput and low-cost approach for the fabrication of nanostructures[1]. The self-assembly is a spontaneous ordered arrangement of the entities. The existence of self-assemblies is in nature itself, viz., proteins, nucleic acids, cell membranes, etc.; thus, self-assemblies are common in biology[2]. Chemistry has a dedicated branch called supramolecular chemistry, where self assemblies are studied using the molecule as a super-atom. Such nanoparticle super-atom based self-assemblies have been studied in different dimensional, 3D, 2D (slab or monolayers), and 1D (wire) forms [1, 3-5]. The self assemblies mainly have applications in optoelectronics, transport, drug and gene delivery, etc.[2 6].

Apart from the molecules, atomic clusters are also promising candidates for the nanoparticle assembled materials[1 7 8]. Clusters as superatoms give us an handle to create unique compounds with tunable properties[9]. It is also reported that specific clusters mimic the properties of certain groups in the periodic table[10]. Most cluster assemblies stabilize in the *FCC* or hexagonal lattice [11 12]. It is possible to produce a large number of size selected cluster assembled nanostructures of specific sizes by using chemical methods like chemical solution deposition[13] or physical methods like cluster beam technology[14]. However, clusters may form the coalesce resulting in the loss of intactness and identity of cluster during the assembly synthesis[15]. The issue of coalescing of the clusters can be avoided by passivation using suitable ligands; however, this may affect the intrinsic properties of the cluster. Therefore, there has been a constant search for stable atomic clusters for self-assembly. The prediction of stable clusters is considerably assisted by the magic numbers, especially for the metallic clusters[16]. However, such magic numbers are not yet developed for the covalently bonded clusters[17].

The II-VI group compound semiconductors are crucial in material science as they have applications in the photo-voltaic cells[18 19], radiation detectors[20] and spintronics[21 22]. The clusters of II-VI semiconductors are among those studied for self-assembly since the early days of the field[23]. Saraf et al.[24] discuss the tunability of properties of self-assembled $Cd_6Se_6$ clusters upon doping of transition metal atoms. The self-assemblies with CdTe nanoparticle as super-atom in different dimensions has been reported[4 25 26]. Cadmium Telluride (CdTe) is a member of the II-VI group, has the zinc blend structure with tetrahedral bonding. It has a direct bandgap of 1.5 eV [27] and, therefore an essential material for solar cell applications. Interestingly, among the II-VI group compounds, CdTe nanoparticles resemble the proteins in their degree of anisotropic interactions in solution, surface chemistry and physical dimensions[28 29].

The fullerene is a well-known cage structure used as a super-atom for the cluster assembled crystals. On the same basis, there has been a constant search for such fullerene-like cage structures to be potential super-atoms. Therefore, it is interesting to explore the cage-like clusters of compound semiconductor CdTe for self-assembled materials. CdTe clusters have different geometries, ranging from wire to spheroid and its smallest stoichiometric cluster with a spheroidal or cage-like form is the $Cd_9Te_9$

[30 31 32]**.** More details on structure and properties of the $Cd_9Te_9$ cluster are discussed in our earlier work [31].

This paper reports the properties of the cluster assembled nanostructures using the $Cd_9Te_9$ cluster as a super-atom. The geometry, dynamical stability, electronic structure, and optical properties have been studied for the various nanostructures formed in 1D, 2D and 3D. We present the results for self-assembled cubic lattices (*SC*, *BCC*, *FCC*, and *ZB*), followed by the five different types of cluster assembled monolayers in 2D. In addition, we also study the cluster assembled wires which correspond to 1D solid.

**Methodology**

The calculations, viz., geometry optimization, *ab-initio* MD, electronic structure, phonon dispersion, and optical properties, are performed using density functional theory as implemented in the VASP code[33 34]. For approximating the exchange-correlations functional, the generalized gradient approximation (GGA) due to Perdew–Burke–Ernzerhof (PBE)[35] was used in the projected augmented wave (PAW) flavor for the pseudopotentials[36]. A plane wave cutoff energy of 600 eV (for 3D) and 700 eV (for 1D and 2D) was used. The stopping criterion for total energy convergence for self-consistency was $10^{-7}$ eV. The conjugate gradient method is employed for quenching structures to the nearest local potential energy minimum [37]. These structures are relaxed until the maximum forces on individual ions are below $10^{-3}$ eV/Å for stopping the ionic relaxation cycles.

The geometry of $Cd_9Te_9$ cluster (fig. 1), the super-atom for cluster assembly, has been obtained from the bulk fragment as an initial input which undergoes the simulated annealing followed by constant temperature MD and final quenching of numerous configurations [31]. It can be visualized as comprising of three rings of $Cd_3Te_3$ connected in parallel [31], $R_1R_2R_1$ arrangement with a central hexagonal cavity.

Calculations have been performed for the different cluster assembled bulk (CAB) in cubic systems, viz., *SC*, *BCC*, *FCC*, and *ZB*, using the $Cd_9Te_9$ cluster as the basis. All the structures are treated as primitive and with only translational symmetry (space group P1). A gamma centered k-point grid of 8×8×8 was used for all the calculations, while a denser k-mesh was used for the density of states (DOS) calculations. The cells are subjected to at least twenty rounds of alternate ion and volume relaxations keeping the shape of the cell fixed. These relaxed CABs are then equilibrated at 300 K to obtain the stable configurations at room temperature using the *ab-initio* molecular dynamics (AIMD). The CAB structures are slowly heated to 300 K using simulated annealing within the micro-canonical ensemble. Then at 300 K, a constant temperature AIMD is performed within a canonical ensemble for more than 10 ps in the steps of 1 fs. The idea was to see if the structures are stable at room temperatures. The equilibrated structures were again subjected to the relaxation procedure as above to obtain the final relaxed configurations at 0 K when the equilibrated configurations differed from the starting relaxed configurations. After the geometry optimization, the electronic structures, phonon dispersions and optical properties were calculated for these CABs. For the frozen phonon calculations, a 2×2×2 super-cell (1×1×1 for ZB) was used along with a k-mesh of 4×4×4 gamma centered grid. The bulk modulus of these structures was calculated by isotropically distorting the volume in two different ways: (i) by treating the clusters as hard particles with fixed atoms, which varied only the bonding distances between the hard clusters to mimic the clusters as pseudo-atoms and (ii) by treating the clusters as soft particles, which varied all the inter-atomic distances.

These CABs (*SC*, *BCC*, *FCC*, and *ZB*) have structures that are formed from the stacked 2D monolayers. Therefore, the cluster assembled monolayers (CAM) are derived from these structures as: (010) layer of *SC* (referred to as *A*), (011) layer of *BCC* (*B*), (111) layer of *FCC* (*C*) and (001) layer of *ZB* (*D*) and oriented in the *x-y* plane with vacuum in the *z*-direction. While the *A*, *B* and *C* monolayers are planar in nature, the *D* monolayer is buckled. Along with these CAMs, another CAM (*E*) has been created by putting an isolated cluster in the *x-y* plane, so that the cluster is free to reorient with the monolayer remaining planar. Similar to the bulk, these monolayers are also subjected to the ionic relaxations and the AIMD runs at 300 K to see their structural stability at higher temperatures. The electronic structure, phonon dispersion and optical properties have been calculated for these monolayers using the energy cutoff of 700 eV for plane waves. The K mesh used is different due to the different sizes of monolayer unit cells and are as follows *A*: 5×5×1, *B*: 5×5×1, *C*: 3×3×1, *D*: 3×3×3 (accounting for the buckled nature of the monolayer) and *E*: 5×5×1. The *A* and *E* monolayers have a single cluster in their unit cell, while the rest of them have two. Further, cluster chain structures can be derived from the CAMs and similar studies have been performed for them too. Three different chains have been studied here, which can be distinguished based on their inter-cluster bondings. The K mesh of 5×1×1 is used for the calculations of cluster assembled chains (CAC).

In all the simulations of monolayers, chains, a sufficiently large vacuum of 35 Å was used in all the required directions to minimize the interactions with their images arising due to periodic boundary conditions.

## Results

The cluster assembled materials are studied using small clusters of CdTe as the building blocks. The $Cd_9Te_9$ is a cage-like cluster with $C_3$ symmetry. The properties of the cluster are discussed in our earlier work [31], while it is used as a super-atom in this work. The geometry of the $Cd_9Te_9$ cluster is shown in fig. 1.

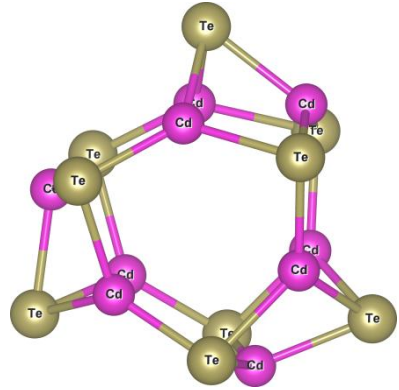

Figure 1: The geometry of $Cd_9Te_9$ cluster. It has a cage-like or spheroidal form with $C_3$ symmetry. The outermost atoms are Te (Olive green).

The cluster assemblies have been studied in different conventional cubic crystal structures: simple, body center, face center and zinc blend. The primitive unit cell of *ZB* has two cluster super-atoms as the basis, while the rest have one.

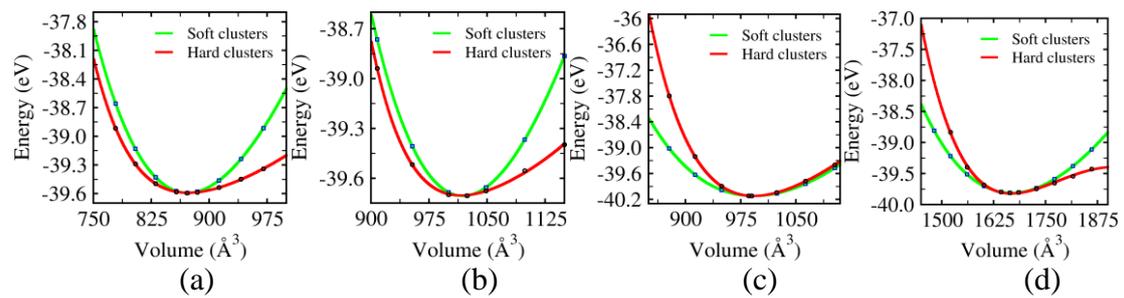

Figure 2: The total energy vs. unit cell volume of CAB in different cubic systems viz., (a) *SC*, (b) *BCC*, (c) *FCC* and (d) *ZB*. The clusters are treated as the soft (green) and hard (red) particles while varying the volume isotropically.

Upon initial relaxations, CABs were observed to have structures like 2D stacks (*SC*, *BCC*, *FCC*) or zig-zag chains (*ZB*). The variations in the total energy versus the cell volume (per formula unit) are shown in fig. 2 for the cases where clusters are treated as soft or hard clusters. The volume of the soft cluster changes with the change in the cell volume, while the volume of hard cluster does not change. In energy vs.

volume calculations, the volume of the cluster (inter-atomic separations) is scaled linearly with the shrinking or expansion of the cell for soft clusters, whereas for hard clusters, only the inter-cluster bond lengths are scaled with the cell volume. The bulk modulus ($B_0$), its pressure dependence ($B_0'$) and volume ($V_0$) obtained from the third order Birch-Murnaghan equation of state are given in table 1.

| Cubic system | Hard cluster | | | Soft cluster | | |
|---|---|---|---|---|---|---|
| | $B_0$ (GPa) | $B_0'$ | $V_0$ (Å$^3$) | $B_0$ (GPa) | $B_0'$ | $V_0$ (Å$^3$) |
| *SC* | 11.64 | 13.73 | 872.898 | 23.98 | 4.72 | 872.14 |
| *BCC* | 9.96 | 17.74 | 1016.131 | 19.99 | 4.70 | 1018.80 |
| *FCC* | 29.59 | 13.38 | 992.95 | 20.90 | 4.63 | 993.03 |
| *ZB* | 10.21 | 23.80 | 1669.75 | 12.26 | 4.64 | 1667.81 |

Table 1. The bulk modulus ($B_0$), its pressure dependence ($B_0'$) and volume ($V_0$) of CAB in different cubic lattices. While distorting the volume isotropically, the cluster as a whole is treated as if the hard and soft clusters, separately.

Then, the stability of these cluster assemblies at room temperature was studied using the *ab-initio* molecular dynamics. The *FCC* and *ZB* CAB systems found themselves in the new, lower energy configurations with structural changes in constant temperature AIMD. Therefore, these low-energy configurations at room temperatures are quenched to the nearest local minima using the conjugate gradient method[37]. The final CAB structures are shown in fig. 3 and the constant temperature MD results are shown in fig. S1 in the supplementary information (SI). The structural changes are reflected as a sudden change in the average total energy in the MD simulations (fig. S1c, S1d). The difference in the average energy at 300 K for the *FCC* is 0.8 eV, whereas, for *ZB*, it is 0.3 eV.

*ZB* CAB's earlier zig-zag chain-like structure settles in the 2D stack form with buckling. The inter-cluster bond length has reduced from 3.02Å to 2.89Å in *FCC*, whereas in *ZB*, it has increased from 2.92Å to 2.97Å. The lattice constant in *FCC* decreases by 0.85% and in *ZB* by 2.5%, while the energy gap increases by ~0.25 eV only in the *FCC* structure. The binding energy with respect to the fragmentation of the nanostructures into $Cd_9Te_9$ clusters has the order *FCC>ZB>BCC>SC*, as shown in table 2.

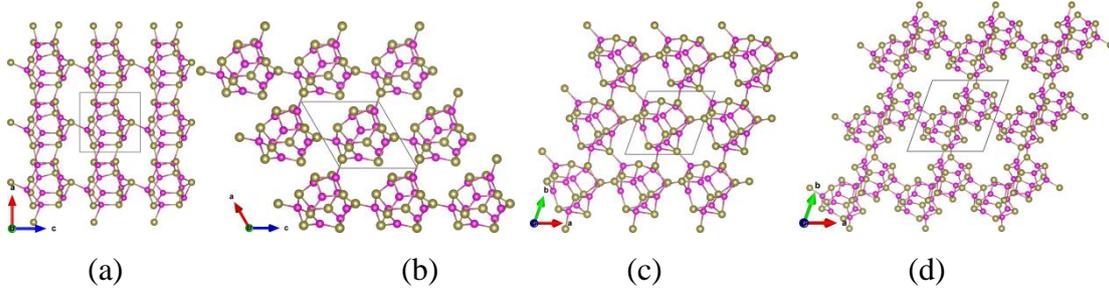

(a)          (b)          (c)          (d)

Figure 3. The geometry of cluster assembly in (a) *SC*, (b) *BCC*, (c) *FCC* and (d) *ZB* cubic structures. The structures are oriented in (010) for *SC*, and *BCC*, (001) for *FCC* and *ZB*.

Fig. 4 shows the phonon dispersions obtained for all the CABs. All the structures have all real frequencies at the zone center (Γ-point) in the Brillouin zone (BZ). As we move towards the zone boundaries, all the optical modes remain real, while one or more acoustic modes become imaginary, although small in magnitude ($< 10$ cm$^{-1}$). In the *FCC* case (fig. 4c), we have doubly degenerate acoustic mode with imaginary frequencies only along the Γ-L direction, while the frequencies are real in other directions. In the *SC*, *BCC* and *ZB* cases (fig.4 a, b and d), some of the acoustic modes have imaginary frequencies in the whole Brillouin zone. A small gap around 75 cm$^{-1}$ can be seen to be present in all the phonon dispersions. The element-wise contribution to these modes can be seen from the corresponding phonon PDOS. The vibrational modes have more contribution due to the *Cd* than *Te* in the 40 cm$^{-1}$ - 75 cm$^{-1}$ and above 150 cm$^{-1}$ frequency ranges, whereas the Te movements contribute more in the frequency range of 80 cm$^{-1}$ to 150 cm$^{-1}$. At low frequencies, optical modes are due to the twisting and breathing modes of clusters in different directions (fig. S2 in SI), where there is an equal contribution from *Cd* and *Te*. Beyond that, there are vibrations dominated by either *Cd* or *Te*. The internal modes are seen at higher frequencies near 170 cm$^{-1}$ with the comparable contribution of *Cd* and *Te*. The bands are seen to have very sharp dispersions and the internal modes are higher in frequency as compared to the CdTe bulk crystal (~145 cm$^{-1}$). This is a consequence of the cluster having much stronger bonds as compared to those in the bulk CdTe crystal and the low dimensionality of the cluster which is used as the super-atom. The phonon dispersions show that the *FCC* nanostructure is the most stable and it also has the largest binding energy as shown in table 2.

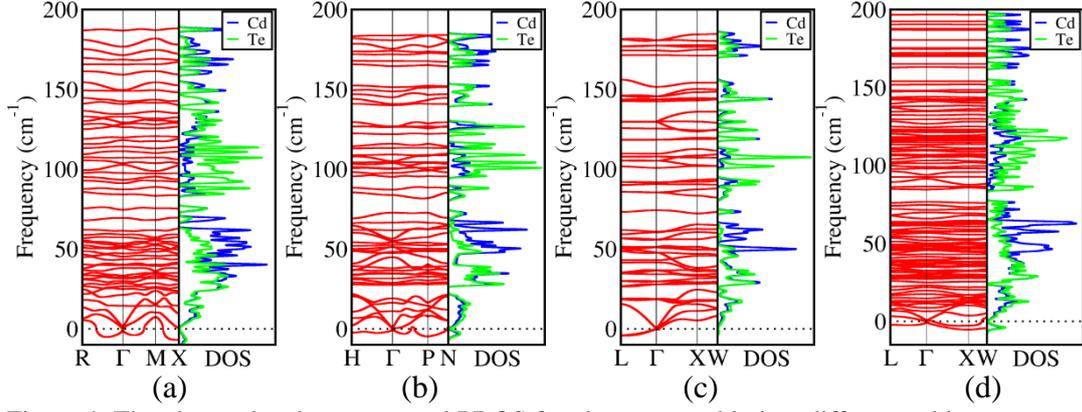

Figure 4: The phonon band structure and PDOS for cluster assembly in a different cubic structure, viz. (a) *SC*, (b) *BCC*, (c) *FCC* and (d) *ZB*.

Electronic band structures and DOS were calculated and the results are shown in fig.5 for all the structures. For reference, the DOS of the building block of CAB, i.e., isolated cluster $Cd_9Te_9$, is also shown in this figure. The isolated cluster has a band gap of 1.6 eV. The *BCC* and *ZB* CAB have a direct bandgap of 1.4 eV and 1.78 eV respectively, whereas the *SC* and *FCC* have an indirect bandgap of 1.0 eV (direct gap 1.3 eV) and 1.56 eV (direct gap 1.66 eV), respectively. The Fermi energies of these CAB are as follows: -0.0136 eV (*SC*), -0.8024 eV (*BCC*), -0.7770 eV (*FCC*) and -2.6729 eV (*ZB*).

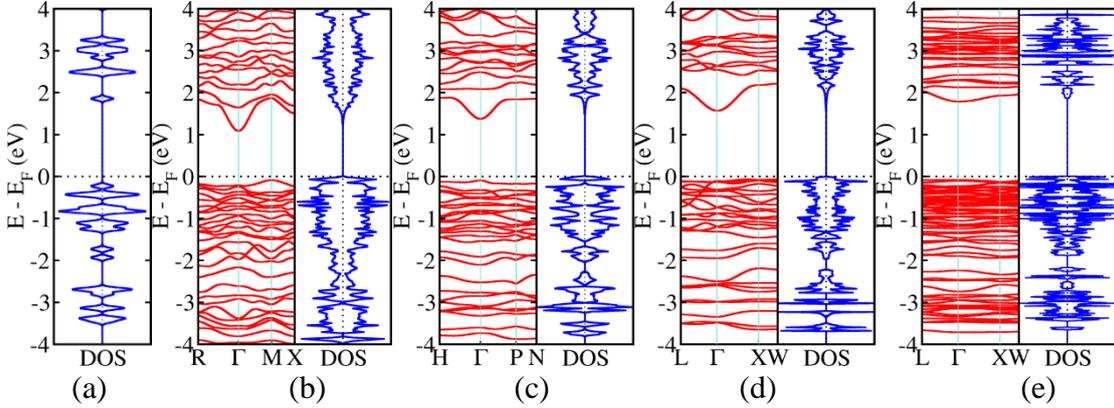

Figure 5: The electronic band structure and density of states for cluster assembly in different cubic structures viz. (a) $Cd_9Te_9$ cluster (for reference), (b) *SC*, (c) *BCC*, (d) *FCC* and (e) *ZB*.

The calculated optical constants for the nanostructures are shown in fig. 6, which shows the real and imaginary parts of the dielectric function and the absorption spectrum in the 0 - 10 eV energy range shown in fig. S3. The magnitude of the static dielectric constant ($\varepsilon_0$) is maximum for *SC* and minimum for the *ZB* case, showing that the *SC* structure is most polarizable. All the nanostructures show maximum absorption in the 6 - 7 eV energy range, dominated by the inter-band absorptions, even though the maximum in the imaginary part of the dielectric constant is ~4 eV in all the cases. Thus all the nanostructures have good absorption in the UV range of the optical spectrum.

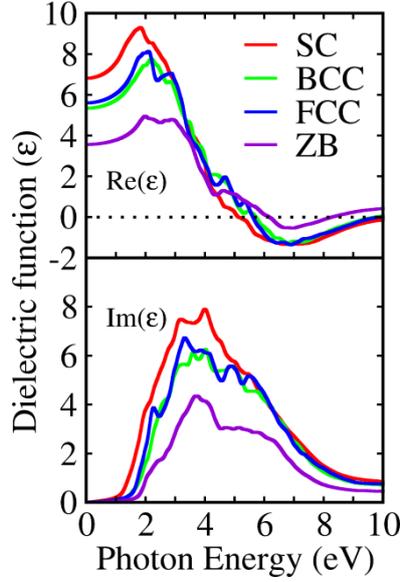

Figure 6: The real and imaginary part of dielectric function (ε) as a function of incident photon energy (eV) for cluster assembly in different cubic systems (*SC*, *BCC*, *FCC* and *ZB*).

Since all CAB systems are formed from stacked 2D layers, we can study the properties of these 2D isolated monolayers. Various such CAMs are derived from the CAB viz., *SC*(010), *BCC*(011), *FCC*(111), and *ZB*(001). Apart from these, a monolayer not derived from the stacked layers is also studied. The top views of their geometries are shown in fig. 7. We refer to these five monolayers as *A*, *B*, *C*, *D*, and *E*. The number of bonds the clusters have with their neighbors in different monolayers is *A*(6), *B*(4), *C*(6), *D*(6) and *E*(4). The binding energies of these CAM have the following order; *C*>*D*>*B*>*A*>*E*.

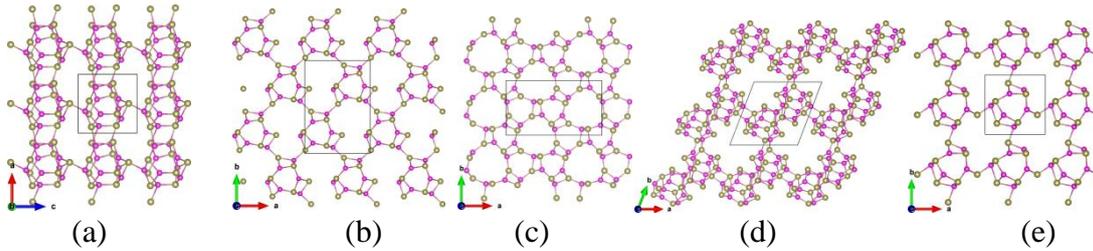

   (a)            (b)            (c)            (d)            (e)

Figure 7. The geometry of CAM derived from bulk CAB (a) *A*, (b) *B*, (c) *C*, (d) *D* and one by confining isolated CdTe cluster in one direction, i.e., (e) *E*.

Unlike the bulk cases where the structure undergoes considerable changes in the course of MD at room temperature, the initial geometries of monolayers don't show any significant deviations from the relaxed geometries at 0 K in AIMD. The total energy vs. time of MD simulations for monolayers are shown in fig. S4. The phonon dispersions and atom projected phonon DOS are shown in fig. 8. Like the CAB, CAMs also have all real frequencies at the zone center of BZ and one or more acoustic modes with imaginary frequencies (< 10cm$^{-1}$) as one goes towards the zone boundary. The monolayers *A* and *C* have one acoustic mode with imaginary frequencies (< 3 cm$^{-1}$).

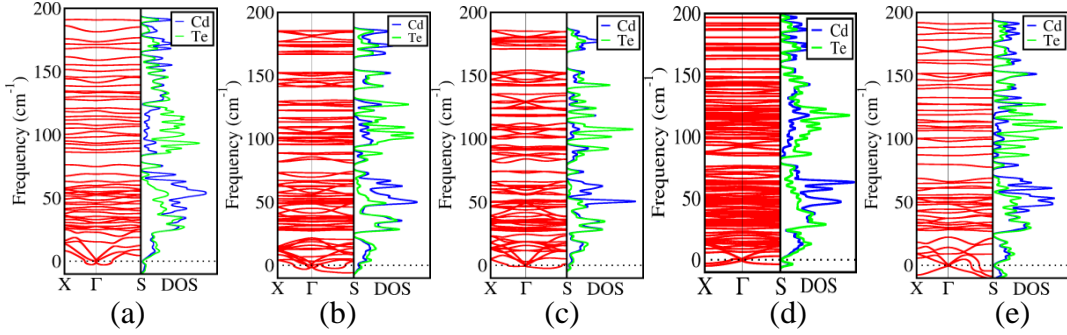

Figure 8: The phonon band structure and partial density of states for CAM derived from different cubic structures, (a)*A*, (b)*B*, (c)*C*, (d)*D* and(e)*E*.

In addition to 75 cm$^{-1}$ as in bulk, a small gap can be seen at 160 cm$^{-1}$ in all the phonon dispersions of monolayers. The element-wise contribution to these modes can be seen from the phonon PDOS and are consistent with CAB. The phonon dispersions show that the monolayers *C* followed by *A* are the most stable. The monolayer *C* also has the largest binding energy, as shown in table 2. The sharp dispersions and internal modes at higher frequencies than bulk CdTe crystal reveal that the bonds in CAMs are stronger.

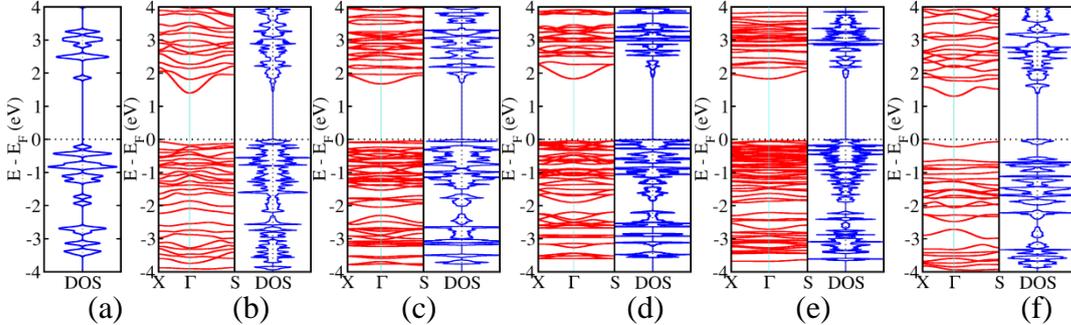

Figure 9: The electronic band structure and density of states for different CAMs,(a) isolated cluster, monolayers (b) *A*, (c) *B*, (d) *C*, (e) *D* and (f) *E*.

The electronic band structure and DOS of CAM are shown in fig. 9. The DOS of the bare cluster is given for reference. A couple of monolayers have the indirect bandgap, *A* (1.4 eV) and *E* (1.3 eV) and the rest have a direct bandgap. The direct band gaps for monolayers are as follows: *A* (1.6 eV), *B* (1.7 eV), *C* (1.8 eV), *D* (1.8 eV) and *E* (1.5 eV). The Fermi energies of these monolayers are given in table 2 and have the order as follows: *E*>*A*>*B*>*C*>*D*. The calculated dielectric function is shown in fig. 10 and absorption spectra in fig. S5. The dielectric constants for the monolayers in-plane and out-of-plane directions are shown in figs. 10(a) and (b). The dielectric behavior doesn't differ significantly in the two directions. The low frequency dielectric constant is maximum for monolayer *A* followed by *E*, whereas for monolayer *D*, it is minimum. It shows that the polarizability is higher for the monolayer *A* and *E*. Unlike the CAB, the peak in dielectric constant and absorption spectra coincides in energy, 4 eV (fig. S5).

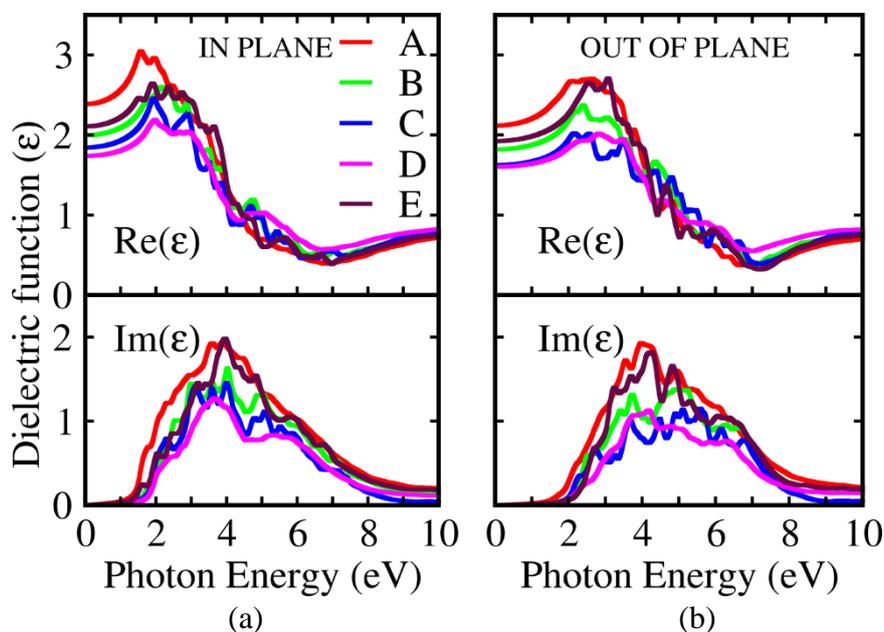

Figure 10: The real and imaginary part of the dielectric function for different CAMs (a) in-plane and (b) out-of-plane directions.

Three different cluster chains are studied based on inter-cluster connections in CAM. The geometries of CACs are shown in fig. 11 and are marked as *I* (fig.11a), *II* (fig.11b) and *III* (fig.11c). The inter-cluster bonds that each chain has are 4 (*I*), 1 (*II*) and 2 (*III*).

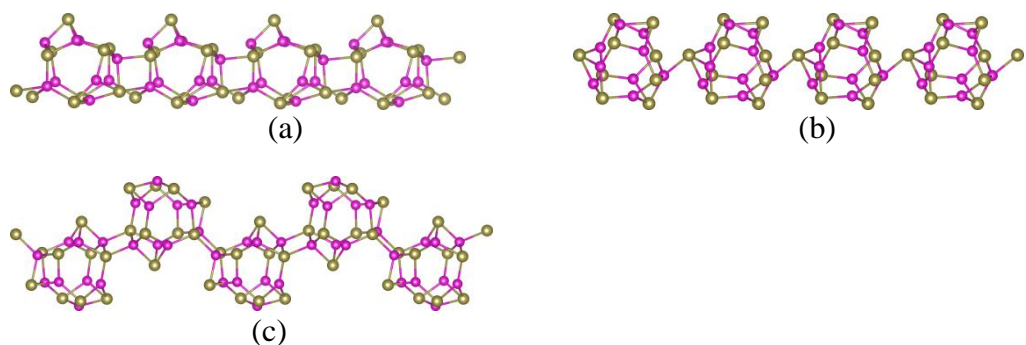

Figure 11: Geometries of CACs, (a) *I*, (b) *II* and (c) *III*.

These chains are studied for stability at room temperature and the energy fluctuations from AIMD equilibration are shown in fig. S6. Like CAB *FCC* or *ZB*, the *I* chain also has relaxed further to lower energy configuration with minor structural changes in AIMD at 300 K. One of the inter-cluster bond lengths for the chain *I* have reduced from 3.26 Å to 2.85 Å. Therefore, this new configuration has been used for further calculations after relaxing to the nearest local minima at 0 K. The other two chains (*II* and *III*) are stable at 300 K with an intact initial structure. Apart from the stability at room temperature, phonon dispersions are carried out for these chains (fig. 12). Chains *I* and *II* have all real frequencies at the BZ center, whereas chain *III* has imaginary acoustic and optical frequencies. Therefore chain *III* is structurally unstable, which might arise from its zig-zag structure. The one or two acoustic modes with a small magnitude of imaginary frequencies are observed for the chain *I* and *II* away from the zone center.

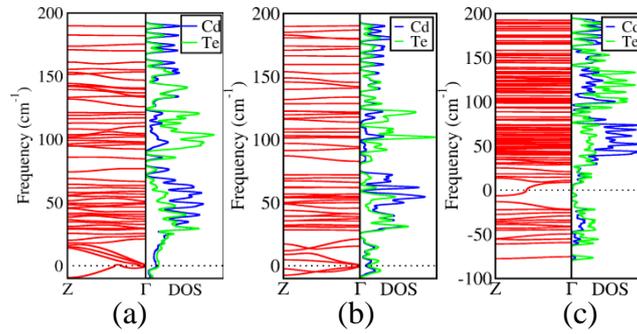

Figure 12: The phonon band structure with corresponding DOS for CACs, (a) *I*, (b) *II*, (c) *III*.

The inter-cluster bond lengths are as follows, chain *I* (2.85 Å- 3.04 Å), chain *II* (2.93 Å) and chain *III* (2.92 Å). The electronic structures (fig. 13) suggest that chain *I* has the indirect bandgap of 1.65 eV and the other two have direct band gaps (*II* - 1.6 eV, *III* - 2.0 eV). The binding energy is largest for the chain *I* and their order are *I* > *III* > *II*.

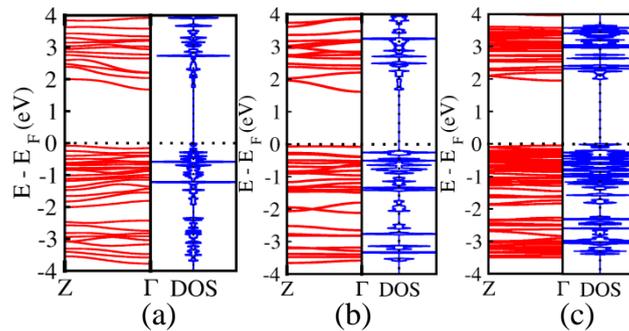

Figure 13: The electronic band structure with corresponding DOS for CACs, (a) *I*, (b) *II*, (c) *III*.

The optical properties of cluster chains have been studied for the incident photon energy of 0 eV to 10 eV (fig. 14 and fig. S7). The dielectric behaviour is calculated in two different directions, viz., along the axis and normal to the axis of the chain. In the former case, chain *I* has a higher dielectric constant, whereas in the latter case, it is chain *II*. Dielectric properties of chain *III* do not depend on the direction of the incident photon. The absorption spectrum is calculated for the incident photon along the axis of the chain. The absorption is more for the chain *I* up to ~4 eV and for chain *II* beyond that. Chain *III* has the least absorption throughout the given energy range.

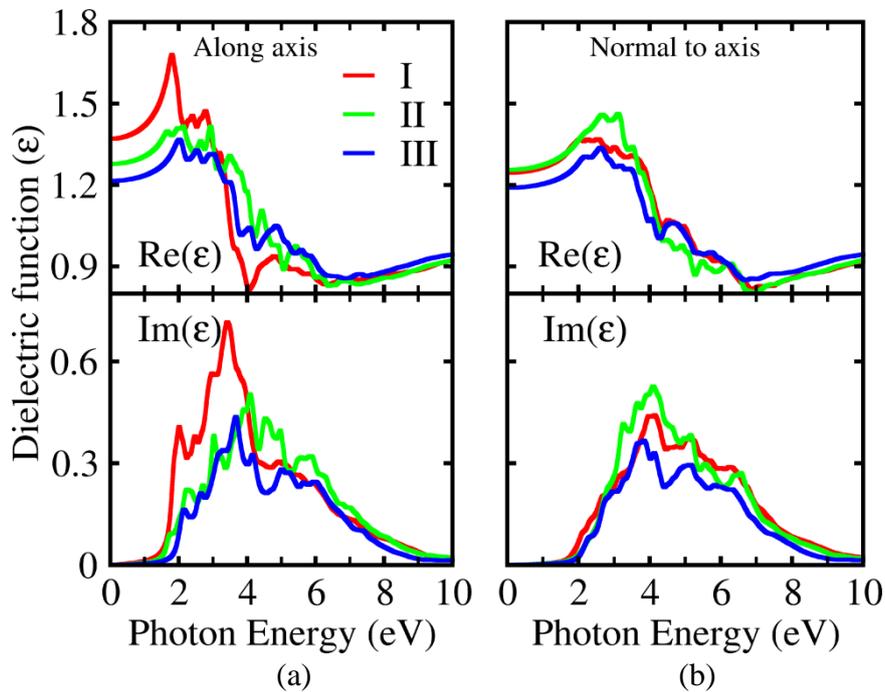

Figure 14: The real and imaginary part of the dielectric function (a) along-chain axis and (b) perpendicular to the chain axis of different CACs.

## Discussion

The templated cluster assembly has been studied in the conventional cubic crystal structures as the simplest model for agglomerating the clusters or nanoparticles. These cubic systems can be characterized based on the lattice constant, packing fraction, binding energy, bandgap, etc. The packing fractions, which are calculated using void fraction of respective conventional unit cells for each CAB are *SC* (41.7%), *BCC* (35.6%), *FCC* (36.3%) and *ZB* (21.7%). Overall the lower packing than the bulk CdTe is due to the 2D stack-like structures of CAB which are bonded via the van der Walls interactions across the layers. The distance between the 2D stacks is maximum for the *ZB* CAB (6.30 Å) and minimum for *SC* CAB (2.86 Å). As a result of the low packing fraction in the *ZB* case, it has the smallest bulk modulus and is highly dependent on pressure. The bulk modulus is largest for the case of *FCC* in both cases. The inter-cluster bond lengths are between 2.86-3.32 Å for *SC*, 2.93-2.95 Å for *BCC*, 2.89 Å for *FCC* and 2.79-2.97 Å for *ZB*. Thus the *FCC* structure has strongest inter-cluster bonds. The direct band gaps range from 1.3 eV to 1.78 eV, with a couple of CAB having the indirect band gaps. The phonon dispersions and AIMD show that these conventional cubic structured cluster assemblies are stable under ambient conditions. Such assemblies can be useful for solar cell applications due to their appropriate band gaps.

The CAMs differ from each other in the bondings and the cluster orientations. The monolayers have different thicknesses, 7.069 Å (*A*), 5.2876 Å (*B*), 5.153 Å (*C*), 7.524 Å (*D*), 5.405 Å (*E*). The clusters are oriented such that the long axis is out of the plane in the *A* monolayer, while it is in-plane in other monolayers. The *D* monolayer has the highest thickness since the clusters are not in-plane and are arranged in a zig-zag manner. The inter-cluster bond lengths for the monolayers are, 2.86 Å - 3.19 Å (*A*); 2.92 Å - 2.95 Å (*B*); 2.90 Å (*C*); 2.78 Å - 2.97 Å (*D*) and 2.82 Å - 2.91 Å (*E*). These inter-cluster bond lengths are consistent with the CAB. The CAMs studied here are stable dynamically and *A* and *C* monolayers are the most stable. The electronic structure suggests that the *A* (1.40 eV) and *E* (1.30 eV) have the indirect bandgap, whereas the rest have a direct energy bandgap. As expected, the energy gaps of CAMs are higher than their corresponding CABs. Although monolayer *C* is a derivative of *FCC* CAB, unlike the bulk, it has direct bandgap, implying that the stacking leads to an indirect gap in this case.

| Self Assembly | | Binding energy (eV) | $E_g$ (eV) | $\Delta E_F$ (eV) |
|---|---|---|---|---|
| **Isolated cluster** | | - | 1.60 | 0 |
| **CAB** | *SC* | 0.7156 | 1.30 (1.00) | 4.7177 |
| | *BCC* | 0.8285 | 1.40 | 3.9289 |
| | *FCC* | 1.244 | 1.66 (1.56) | 3.9543 |
| | *ZB* | 0.933 | 1.78 | 2.0584 |
| **CAM** | *A* | 0.5813 | 1.60 (1.40) | 0.832 |
| | *B* | 0.756 | 1.70 | 0.4698 |
| | *C* | 1.243 | 1.80 | 0.1088 |
| | *D* | 0.9374 | 1.80 | 0.0971 |
| | *E* | 0.3787 | 1.50 (1.30) | 1.0748 |
| **CAC** | *I* | 0.7579 | 1.80 (1.65) | −0.3132 |
| | *II* | 0.1098 | 1.60 | −0.1301 |
| | *III* | 0.6297 | 2.00 | −0.4048 |

Table 2. The binding energy with respect to fragmentation into isolated clusters, direct energy bandgap and difference in the Fermi level (with respect to that of isolated $Cd_9Te_9$ cluster) for different self assemblies. Wherever the system has an indirect bandgap, it is mentioned in the bracket. Fermi level of isolated cluster is -4.7313 eV.

Among the CACs, chain *I* has indirect bandgap (1.65 eV) and the rest have direct gap. The chain *III* has the maximum bandgap (2.00 eV). The binding energy and phonon dispersions of chains suggest that chain *I* is most stable. All these self assemblies of clusters (bulk, monolayer and chain) are stable at room temperature, including its isolated building block, the $Cd_9Te_9$ cluster. Finally, the binding energy, bandgap and Fermi levels of all of the cluster assemblies are shown in table 2. A couple of CAB and CAMs are reported to have the indirect bandgap differing by 0.1 to 0.3 eV from the corresponding direct bandgap. The CAB *FCC* and corresponding CAM (*C*) have the highest binding energy in CABs and CAMs, respectively.

## Conclusion

The geometry relaxation, dynamic stability, electronic structure and optical properties have been carried out for the dimension-controlled (templated) static self-assemblies with the smallest cage-like cluster of $Cd_9Te_9$ as the super-atom. The lattices studied include 3D bulk (*SC*, *BCC*, *FCC* and *ZB*), 2D (five distinct self assembled monolayers), 1D wire (three different chains). The 3D CABs form into 2D stacked structures. These structures are stable and semiconducting, with the largest bandgap for the *ZB* structure. The *ZB* structure has largest interlayer separation. The *SC* and *FCC* systems have an indirect bandgap, whereas the *BCC* and *ZB* have a direct bandgap. Further, the *FCC* lattice structure is the most stable based on energetics and dynamic stability studies. The dielectric constant and absorption spectra for *BCC* and *FCC* more or less overlap and lie intermediate to the *SC* and *ZB*. The 2D CAMs are semiconducting with energy gaps ranging from 1.5 eV to 1.8 eV. The monolayer *C* is dynamically more stable and has the highest energy bandgap amongst 2D structures. The in-plane and out-of-plane dielectric constants do not differ much. Among the CACs, *I* and *II* are stable, whereas chain *III* is unstable as there are unstable vibrational modes at the zone center. As expected, from 3D to 2D to 1D, the energy bandgap has increased due to confinement. The energy bandgap, dynamic stability of these self assemblies suggests their potential applicability in photo-voltaic, solar cell, devices.

## Acknowledgment

KTC acknowledges DAE/IGCAR for the research fellowship.

# Supplementary Information

Structural stability, electronic structure and optical properties of dimension controlled self-assembled structures from clusters of cadmium telluride.

Kashinath T Chavan, Sharat Chandra
*Materials Science Group, Indira Gandhi Centre for Atomic Research, Kalpakkam 603102, Tamil Nadu, India*
*Homi Bhabha National Institute, Training School Complex, Anushakti Nagar, Mumbai, 400094, India*

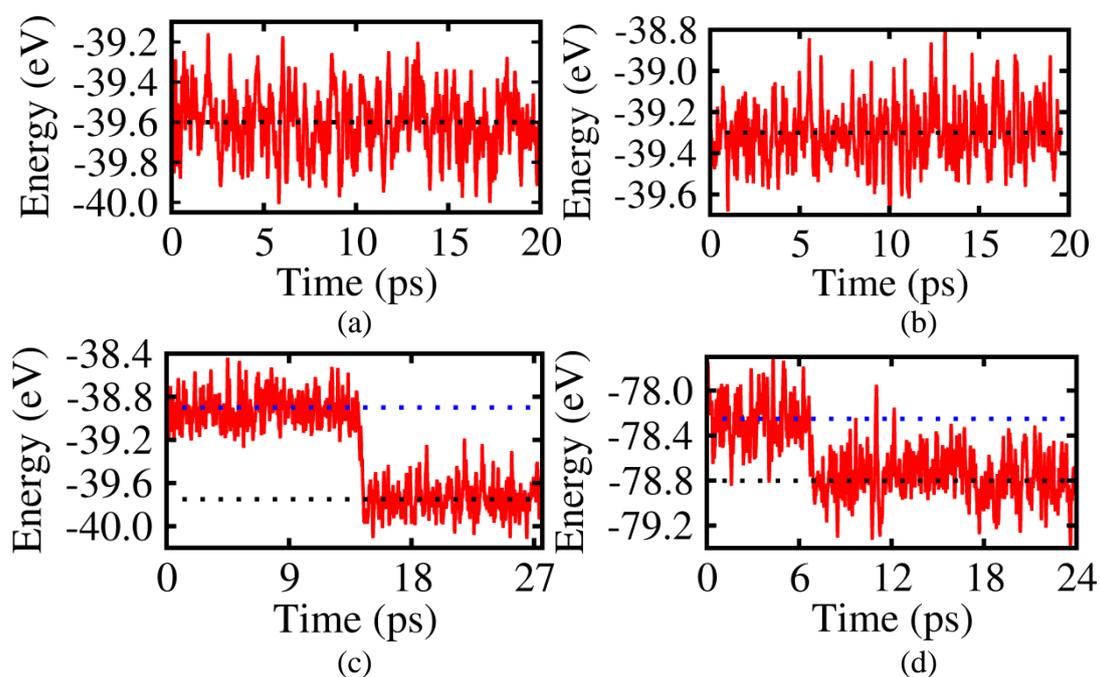

Figure S1: The total energy vs. time step of AIMD for CAB in (a) *SC*, (b) *BCC*, (c) *FCC* and (d) *ZB* at 300 K.

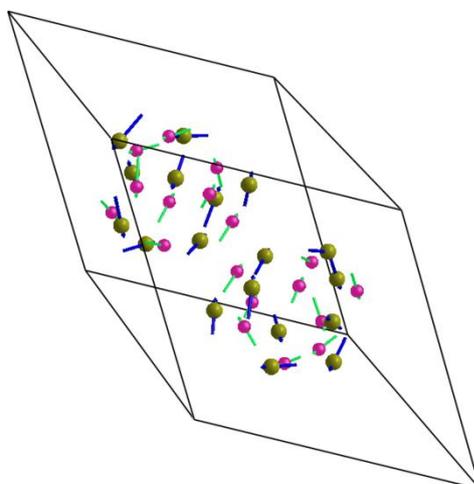

Figure S2. One of the vibration modes for the CAB for the initial optical frequencies which corresponds to the twisting of clusters.

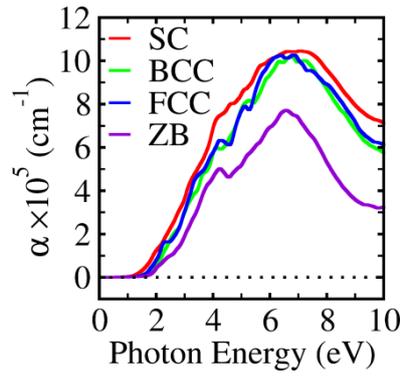

Fig S3: The absorption coefficient (α) as a function of incident photon energy (eV) for cluster assembly in different cubic systems (*SC*, *BCC*, *FCC* and *ZB*).

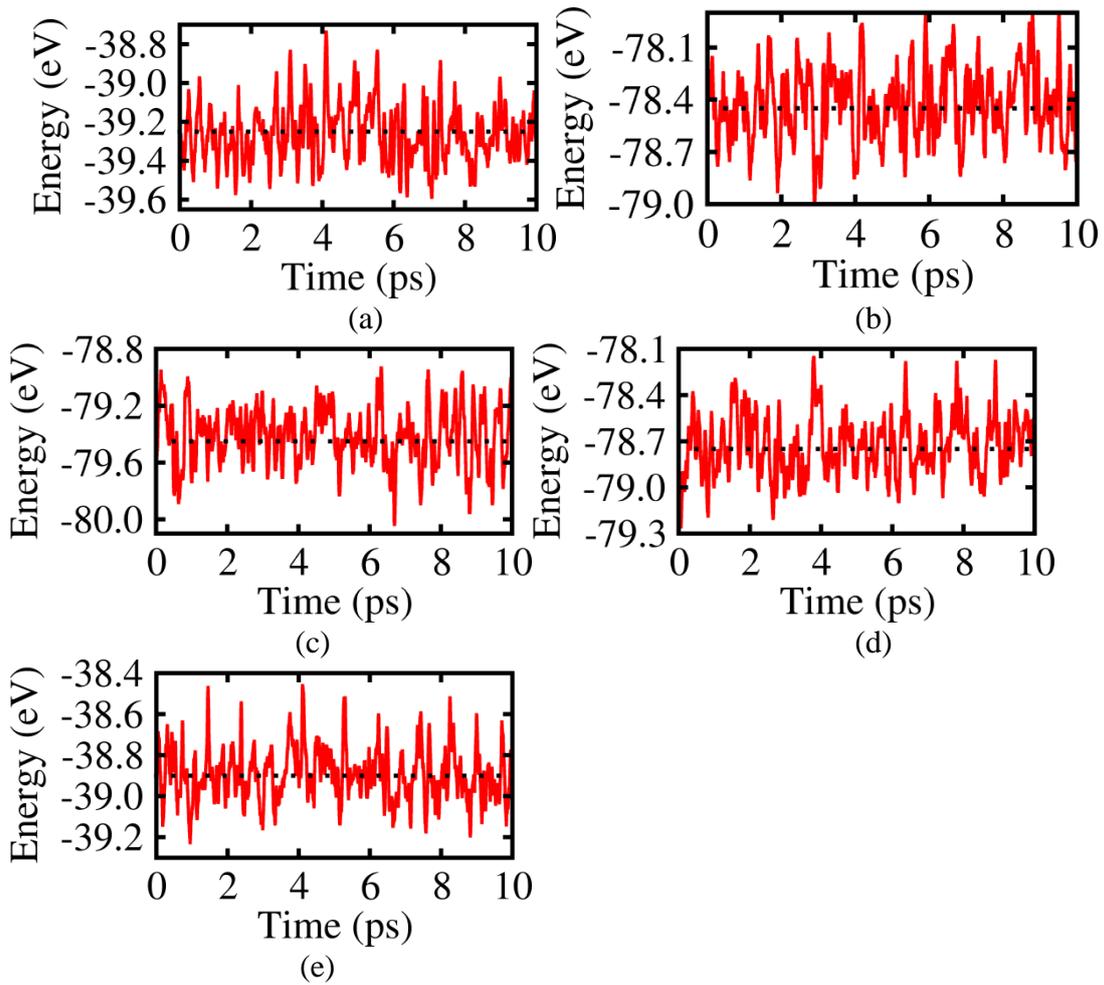

Figure S4: The total energy vs. time steps of AIMD for different CAMs; (a) *A*, (b) *B*, (c) *C*, (d) *D*, (e) *E* at 300 K.

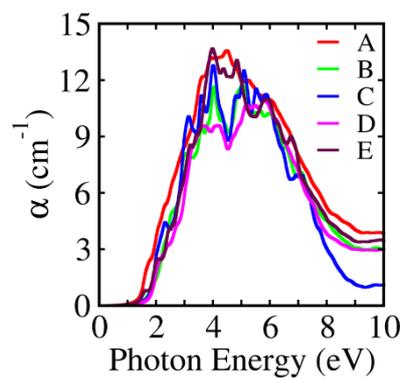
Figure S5: The absorption coefficient for incident photon energy for different CAM.

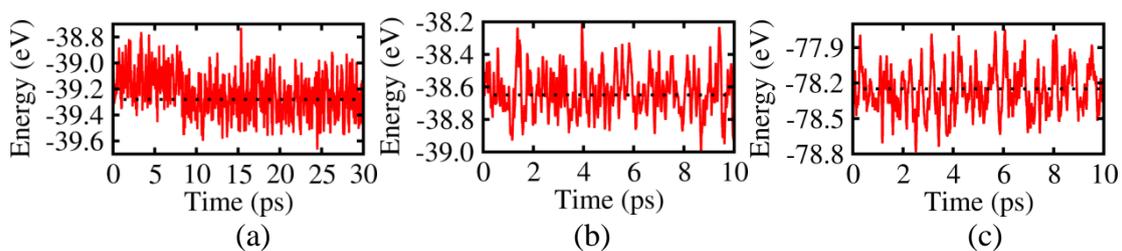
(a)          (b)          (c)

Figure S6: The total energy vs. time steps of AIMD for CAC, (a)*I*, (b)*II*, (c)*III* at 300 K.

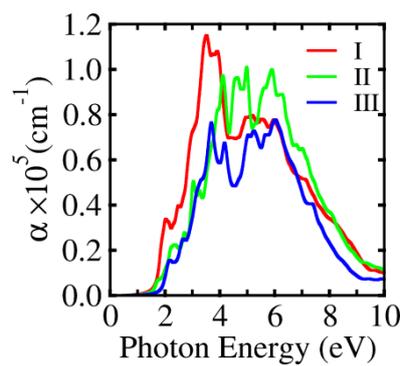
Figure S7: The absorption coefficient for incident photon along the direction of axis for different CACs.